\documentclass[aps,prb,secnumarabic,nobibnotes,twocolumn,superscriptaddress]{revtex4-2}
\usepackage{dsfont}
\usepackage{fontenc}
\usepackage{verbatim}
\usepackage{amsfonts}
\usepackage{mathrsfs}
\usepackage{amsmath}
\usepackage{color}
\usepackage{natbib}
\usepackage{graphicx}
\usepackage{bm}
\usepackage{amssymb}
\usepackage{xspace}
\usepackage{epstopdf}
\usepackage{dcolumn}
\usepackage{multirow}
\usepackage{tabularx}
\usepackage{bbding}
\usepackage{booktabs}
\usepackage[colorlinks=true, letterpaper=true, pdfstartview=FitV, linkcolor=blue, citecolor=blue, urlcolor=blue]{hyperref}

\begin{document}

\title{Essentially degenerate hidden nodal lines in two-dimensional magnetic layer groups}

\author{Xiao-Ping Li}
\email{xpli@imu.edu.cn}
\address{School of Physical Science and Technology, Inner Mongolia University, Hohhot 010021, China}

\author{Chaoxi Cui}
\address{Centre for Quantum Physics, Key Laboratory of Advanced Optoelectronic Quantum Architecture and Measurement (MOE), School of Physics, Beijing Institute of Technology, Beijing 100081, China}
\address{Beijing Key Lab of Nanophotonics \& Ultrafine Optoelectronic Systems, School of Physics, Beijing Institute of Technology, Beijing 100081, China}

\author{Lei Wang}
\address{School of Physical Science and Technology, Inner Mongolia University, Hohhot 010021, China}

\author{Weikang Wu}
\address{Key Laboratory for Liquid-Solid Structural Evolution and Processing of Materials, Ministry of Education, Shandong University, Jinan 250061, China}

\author{Zeying Zhang}
\address{College of Mathematics and Physics, Beijing University of Chemical Technology, Beijing 100029, China}

\author{Zhi-Ming Yu}
\address{Centre for Quantum Physics, Key Laboratory of Advanced Optoelectronic Quantum Architecture and Measurement (MOE), School of Physics, Beijing Institute of Technology, Beijing 100081, China}
\address{Beijing Key Lab of Nanophotonics \& Ultrafine Optoelectronic Systems, School of Physics, Beijing Institute of Technology, Beijing 100081, China}

\author{Yugui Yao}
\email{ygyao@bit.edu.cn}
\address{Centre for Quantum Physics, Key Laboratory of Advanced Optoelectronic Quantum Architecture and Measurement (MOE), School of Physics, Beijing Institute of Technology, Beijing 100081, China}
\address{Beijing Key Lab of Nanophotonics \& Ultrafine Optoelectronic Systems, School of Physics, Beijing Institute of Technology, Beijing 100081, China}

\begin{abstract}
According to the theory of group representations, the types of band degeneracy can be divided into accidental degeneracy and essential degeneracy. The essentially degenerate nodal lines (NLs) are typically resided on the high-symmetry lines of the Brillouin zone.  
Here, we propose a type of NL in two dimension that is essentially degenerate but is hidden within the high-symmetry planes, making it less observable, dubbed a hidden-essential nodal line (HENL). The existence of HENL is guaranteed as long as the system hosts a horizontal glide-mirror symmetry, hence such NLs can be widely found in both non-magnetic and magnetic systems. We perform an exhaustive search over all 528 magnetic layer groups (MLGs) for HENL that can be enforced by glide-mirror symmetry with both spinless and spinfull systems. We find that 122 candidate MLGs host spinless HENL, while 63 candidate MLGs demonstrate spinful HENL. In addition, we reveal that  horizontal mirror and time-reversal symmetry in type-II and type-IV MLGs with spin-orbital coupling can enforce HENL formed. The 15 corresponding candidate MLGs have also been presented. Furthemore, we derive a few typical lattice models to characterize the existence for the HENL. For specific electronic fillings in real materials, namely 4$N$+2 in spinless systems (and 2$N$+1 in spinful systems), the presence of the HENLs in candidate MLGs is required regardless of the details of the systems. Using \emph{ab-initio} calculations, we further identify possible material candidates that realize spinless and spinful HENL. Moreover, spinful HENLs exhibit a novel persistent spin texture wih the characteristic of momentum-independent
spin configuration. Our findings uncover a new type of topological semimetal state and offer an ideal platform to study the related physics of HENLs. 
\end{abstract}

\maketitle

\section{Introduction}\label{intro}
The exploration of topological metals and semimetals, characterized by their unique band degeneracies and the associated physical properties, has attracted significant attention in recent years~\cite{Chiu2016Classification,toposemi2016,bernevig2018recent,weylDirac,yu2022encyclopedia}. The materials exhibit nontrivial degeneracies that may enhance quantum geometry, including Berry curvature and quantum metric, thereby leading to fascinating physical properties~\cite{wang2023quantum, gao2023quantum, feng2024quantum, cui2024electric}.	 These band crossings in solids can be classified into three categories based on the dimension of the degeneracy manifold; \emph{i.e.}, the zero-dimensional (0D) nodal points~\cite{murakami2007phase, young2012dirac, wang2012dirac, wang2013three, weng2015weyl, meng2020ternary, bradlyn2016beyond}, 1D NLs~\cite{burkov2011topological,kim2015dirac,weng2015topological,fang2016topological,13li2016dirac,PhysRevB.99.121106,PhysRevB.95.235138,feng2017experimental}, and 2D nodal surfaces~\cite{zhong2016towards,liang2016node,wu2018nodal,zhang2018nodal}. Among them, the nodal line exhibits rich characteristics in the aspect of topological connections. For instance, within the Brillouin zone (BZ), the NL may manifest as an isolated nodal ring~\cite{PhysRevLett.113.046401,PhysRevLett.115.026403}, a pair of nodal lines traversing the BZ~\cite{chen2015nanostructured,PhysRevB.96.081106}, or form more complex geometric structures, such as nodal chains, nodal nets, a nodal box, crossed nodal rings, and Hopf links \cite{bzduvsek2016nodal,wang2017hourglass,yu2017nodal, fu2018hourglasslike,sheng2017d,yu2015topological,14kobayashi2017crossing,chen2017topological,yan2017nodal,chang2017topological,chang2017weyl}. According to the tilt effect of the band dispersion, NLs can be further classified into type-I, type-II, and hybrid types~\cite{PhysRevB.96.081106,PhysRevB.97.125143}. Besides, NLs can also be classified into $\pi$ Berry phase NL and $Z_{2}$ monopole charge NL based on the classification of real topology~\cite{PhysRevB.92.081201, PhysRevB.89.235127, PhysRevLett.116.156402, PhysRevB.96.155105, PhysRevX.8.031069}.

\begin{figure}[t]
\includegraphics[width=8.4cm]{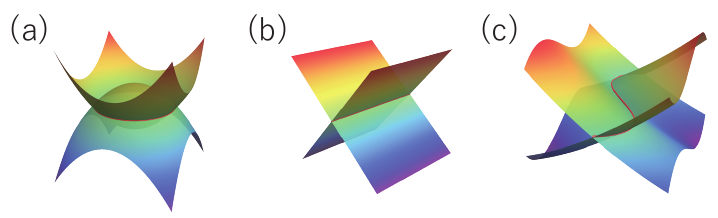}
\caption{Schematic showing the electronic band of nodal line. The red solid line denotes the nodal lines. (a) Accidental nodal line on a high-symmetry plane. (b) The essential nodal line resides on a high-symmetry line. (c) The hidden-essential nodal lines are distributed on a high-symmetry plane.
\label{fig1}}
\end{figure}

Depending on the formation mechanism of band degeneracies, the various types of NLs may be classified into two categories: accidental degeneracies and essential degeneracies. NLs with accidental degeneracies are formed by band inversions in certain regions of the BZ, and lying in high-symmetry planes, as schematically shown in Fig.~\ref{fig1}(a). They can be moved in the BZ or even removed when a symmetry-preserving perturbation is applied. Such as Ca$_{3}$P$_{2}$~\cite{PhysRevB.93.205132}, CaAsP~\cite{yamakage2016line} and most proposed nodal-line materials~\cite{yu2017topological,yang2018symmetry}. The second category of NLs consists of essential degeneracies, which originate from high-dimensional irreducible representations (IRRs) of the little co-group. Consequently, such essential NLs typically reside on high-symmetry lines of the BZ [see Figs~\ref{fig1}(b)]. The traditional essential NLs cannot be moved, let alone removed, as long as the symmetry of the system is preserved. However, when considering the connection of energy bands, essential nodal lines with symmetry-enforced features in nonsymmorphic crystals may emerge. Some of them exhibit movable characteristics, such as off-centered symmetry-protected NLs~\cite{offcentered}, irregular Kramers NLs~\cite{xie2021kramers}, and other types of movable NLs~\cite{PhysRevMaterials.5.054202, PhysRevMaterials.5.124202, figgemeier2024tomographic, PhysRevB.94.195109}. A well-known example of this is the 3D hourglass NLs~\cite{PhysRevB.96.155206}. The hourglass-type band represents a movable essential degeneracy that cannot be directly inferred from the IRRs of the corresponding little group~\cite{PhysRevLett.115.126803}, but originates from nonsymmorphic eigenvalue winding~\cite{wang2016hourglass}. If we extend movable essential degeneracy from 3D to 2D cases, beyond the types already present in three dimensions, will there emerge new movable essential nodal lines? If so, what symmetries enforce them?

In this paper, we address the above question by introducing a type of NLs in two dimensions that is essentially degenerate but emerges on high-symmetry planes (HSPLs). Owing to the presence of this essential NL on HSPLs, its geometric morphology is greatly dependent on the material parameters, resulting in a certain degree of hiddenness [see Fig~\ref{fig1}(c)]. We term such type of nodal line the hidden-essential nodal line (HENL), to distinguish it from the essential NL that resided in high-symmetry line and possess distinct observable characteristics~\cite{PhysRevB.92.081201,PhysRevB.99.121106,PhysRevB.97.045131}. It should be emphasized that, unlike the 2D hourglass nodal line, the existence of HENL does not depend on any high-symmetry points or lines, thereby exhibiting hidden feature. We performed a symmetry analysis and discovered that the presence of a horizontal glide-mirror operation $\left\{M_{z}|\mathbf{\boldsymbol{\tau_{1/2}}}\right\}$ is sufficient to ensure the existence of HENLs in both spinless and spinful systems. And the valence electron filling ensures that the HENLs appears near the Fermi level in the material. Alternatively, a pure horizontal mirror $\mathcal{M}_{z}$ and a time-reversal symmetry $\mathcal{T}$ (or time-reversal symmetry combined with a half lattice translation $\left\{\mathcal{T}|\mathbf{\boldsymbol{\tau_{1/2}}}\right\}$) can also ensure the emergence of HENL in spinful systems. We then develop a strategy to search for them and found that 122 (out of 528) candidate magnetic layer groups (MLGs) can host HENL enforced by $\left\{M_{z}|\mathbf{\boldsymbol{\tau_{1/2}}}\right\}$ in the spinless case, while 63 candidate can host glide-mirror-enforced MLGs in the spinful case owing to the requirement of breaking space-time symmetry $\mathcal{PT}$ to achieve the two-fold NL. Moreover, 15 spinful MLGs are found to host HENL enforced by $\mathcal{M}_{z}$ and $\mathcal{T}$ [or $\left\{\mathcal{T}|\mathbf{\boldsymbol{\tau_{1/2}}}\right\}$] symmetry, specifically in type-II and type-IV MLGs. Then, these results are further confirmed and explicitly demonstrated by the constructed three typical lattice models. Finally, we identify the existence of both spinless and spinful HENL in realistic material candidates, including MoPO$_5$ [without spin-orbit coupling (SOC)] and Sb$_5$O$_{7}$ (with SOC). Interestingly, because of the presence of additional crystalline symmetries, \textquotedblleft free$\textquotedblright$  HENL will be constrained by crystal symmetries, thereby giving rise to intriguing geometric patterns. For example, the HENL in MoPO$_5$ exhibits a fourfold fan-like pattern caused by the $C_{4z}$ rotation symmetry of system, while the HENL in Sb$_5$O$_{7}$ presents a sextuple fan-like pattern because of the rotate-invertion symmetry $S_{3z}$. Furthermore, we note that owing to the mirror symmetry, the spin texture of spinful HENL can be reduced to only the spin $s_{z}$ component, exhibiting the characteristic of spin-momentum locking.
\begin{figure*}[t]
\includegraphics[width=16.8cm]{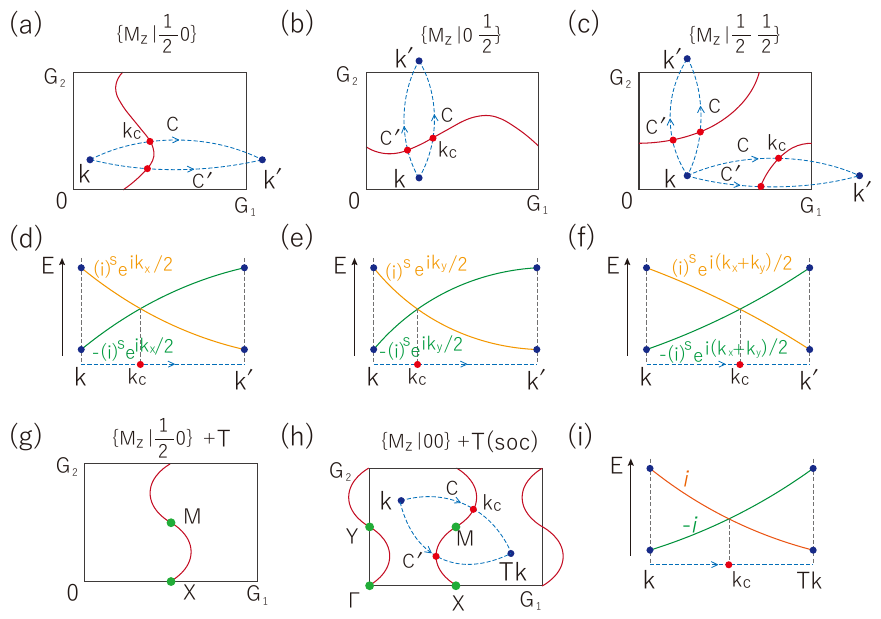}
\caption{Mechanism of emergence of the HENL (a)-(c) Schematic diagrams of the distribution of HENLs enforced by the symmetry of three glide mirrors, where $G_{1(2)}$ denotes the reciprocal lattice vector. (d)-(f) Schematic figure of the energy bands corresponding to (a)-(c) along the curve $C$. The signs $\pm$ of glide eigenvalues are represented by solid lines of different colors (yellow and green). Glide eigenvalues are exchanged
from $\boldsymbol{k}$ to $\boldsymbol{k^{'}}$ point along the path $C$, and a band crossing appears at $\boldsymbol{k}_{c}$. (g) Schematic figure depicting the spinless HENL enforced by $\left\{M_{z}|\frac{1}{2}00\right\}$ with time-reversal symmetry. (h) The spinful HENL guaranteed by ordinary mirror and time-reversal symmetry $\mathcal{T}$ [or $\mathcal{T}\tau_{1/2}$], and its typical energy band along curve $C$ is illustrated in (i). 
\label{fig2}}
\end{figure*}
\begin{table*}[t]
\centering{}%
\caption{The candidate magnetic layer groups that can host spinless HENLs. }
    \begin{tabularx}{\linewidth}{lll}
    \toprule
    \toprule
    Magnetism & Minimal symmetry & \multicolumn{1}{c}{    Magnetic types and candidate magnetic layer groups} \\
    \midrule
    Magnetic & $\left\{ M_{z}|\frac{1}{2}0\right\}$ &  Type-I: \hspace{0.15cm} 5.1.17, 7.1.28, 31.1.188, 33.1.202, 38.1.243, 41.1.276, 43.1.298, 45.1.314 \\
          &       & Type-III: 7.3.30, 31.4.191, 33.4.205, 38.5.247, 38.7.249, 38.9.251, 41.4.279, 41.8.283,  \\
          &       & \hspace{1.3cm} 41.9.284, 43.6.303, 43.7.304,  43.9.306, 45.6.319, 45.7.320, 45.8.321\\
          &       & Type-IV: 4.5.16, 5.4.20, 6.7.27, 7.6.33, 27.9.162, 29.6.179, 31.7.194, 31.9.196, 36.6.226, \\
          &       &  \hspace{1.3cm} 36.9.229, 37.8.237, 37.10.239, 38.12.254, 38.13.255, 40.10.272, 40.12.274,  \\
          &       &  \hspace{1.3cm} 41.11.286, 41.13.288, 48.8.350, 48.9.351, 48.10.352, 48.11.353  \\
          & $\left\{ M_{z}|0\frac{1}{2}\right\}$ & Type-I: \hspace{0.15cm} 28.1.167, 30.1.181 \\
          &       & Type-III: 28.4.170, 30.4.184 \\
          &       & Type-IV: 27.6.159, 27.8.161, 28.7.173, 30.7.187, 36.7.227, 36.8.228 \\
          & $\left\{ M_{z}|\frac{1}{2}\frac{1}{2}\right\}$ & Type-I: \hspace{0.15cm} 32.1.197, 34.1.207, 36.1.221, 39.1.256, 42.1.289, 46.1.323, 48.1.343, 52.1.369,  \\
          &       & \hspace{1.3cm} 62.1.435, 64.1.453 \\
          &       & Type-III: 32.4.200, 34.5.211, 36.5.225, 39.4.259, 39.7.262, 42.5.293, 42.7.295, 42.9.297,  \\
          &       & \hspace{1.3cm} 46.4.326, 46.7.329, 48.4.346, 48.7.349, 52.5.373, 62.4.438, 62.8.442, 62.9.443,  \\
          &       & \hspace{1.3cm} 64.4.456, 64.8.460, 64.9.461\\
          &       & Type-IV: 27.11.164, 28.6.172, 30.6.186, 31.6.193, 31.8.195, 35.7.218, 35.8.219, 37.12.241,  \\
          &       & \hspace{1.3cm} 38.10.252, 38.11.253, 41.10.285, 41.12.287, 47.9.338, 47.11.340, 47.12.341, \\
          &       & \hspace{1.3cm} 51.6.367, 61.10.431, 61.12.433 \\
          \midrule
    Non-magnetic & $\left\{ M_{z}|\frac{1}{2}0\right\}$ &  Type-II: \hspace{0.02cm} 5.2.18, 7.2.29, 31.2.189, 33.2.203, 38.2.244, 41.2.277, 43.2.299, 45.2.315 \\
          & $\left\{ M_{z}|0\frac{1}{2}\right\}$ &  Type-II: \hspace{0.02cm} 28.2.168, 30.2.182 \\
          & $\left\{ M_{z}|\frac{1}{2}\frac{1}{2}\right\}$ &  Type-II: \hspace{0.02cm} 32.2.198, 34.2.208, 36.2.222, 39.2.257, 42.2.290, 46.2.324, 48.2.344, 52.2.370, \\
          &       & \hspace{1.3cm}  62.2.436, 64.2.454 \\
    \bottomrule
    \bottomrule
    \end{tabularx}%
  \label{tab1}%
\end{table*}%
\section{General analysis}\label{general}
We explain the mechanisms of emergence of the HENL in two dimensions. 
Consider a 2D system with a horizontal glide-mirror symmetry $\widetilde{\mathcal{M}}_{z}=\left\{M_{z}|\mathbf{\boldsymbol{\tau_{1/2}}}\right\}$. The $\boldsymbol{\tau_{1/2}}$ is a half of a translation vector, which takes three forms: $\left\{ \frac{1}{2},0\right\}$, $\left\{0,\frac{1}{2}\right\}$ and $\left\{ \frac{1}{2},\frac{1}{2}\right\}$. 
Since $\widetilde{\mathcal{M}}_{z}$ commutes with Hamiltonian $\mathcal{H}(\boldsymbol{k})$ on the glide-invariant plane (that is, the 2D BZ), the related Bloch state can be chosen as the eigenstate of $\widetilde{\mathcal{M}}_{z}$. One finds that $\widetilde{\mathcal{M}}_{z}^{2}=e^{i\boldsymbol{k}\cdot(2\boldsymbol{\tau_{1/2}})}$ in the spinless systems, hence the eigenvalues of $\widetilde{\mathcal{M}}_{z}$ are either $e^{i\boldsymbol{k\cdot\tau}}$ or $-e^{i\boldsymbol{k\cdot\tau}}$, while glide eigenvalues take two values, $ie^{i\boldsymbol{k\cdot\tau}}$ or $-ie^{i\boldsymbol{k\cdot\tau}}$, in the spinful systems because of $\widetilde{\mathcal{M}}_{z}^{2}=-e^{i\boldsymbol{k}\cdot(2\boldsymbol{\tau_{1/2}})}$. We now analyze the band crossings resulting from the evolution of these eigenvalues. For any  point $\boldsymbol{k}$ in the 2D BZ, the pairs of band branches can be labeled by eigenvalues of $(i)^{s}e^{i\boldsymbol{k\cdot\tau}}$ and $-(i)^{s}e^{i\boldsymbol{k\cdot\tau}}$ ($s$ = 0 and 1 correspond to the spinless and spinful case, respectively). Our concern lies in how these energy bands evolve from point $\boldsymbol{k}$ to point $\boldsymbol{k^{'}}$, where the selection of point $\boldsymbol{k^{'}}$ is subject to certain constraints. The difference between the two, $\boldsymbol{q}=\boldsymbol{k^{'}}-\boldsymbol{k}$, must satisfy the following conditions.

(i) The value of $\boldsymbol{q}$ must be an integer multiple of the reciprocal lattice vector, that is,
\begin{equation}\label{c1}
\boldsymbol{q} = n\boldsymbol{G}\quad (\ensuremath{n\in\boldsymbol{Z}}).
\end{equation}
This ensures that $\boldsymbol{k}$ and $\boldsymbol{k^{'}}$ have the same energy.

(ii) The value of $\boldsymbol{q}$ needs to contribute an additional phase of -1 in the glide operation, satisfying the condition $e^{i\boldsymbol{q}\cdot\boldsymbol{\tau_{1/2}}}=-1$, which implies that
\begin{equation}\label{c2}
\boldsymbol{q}\cdot\boldsymbol{\tau_{1/2}} =	\pm\pi+2n\pi \quad (\ensuremath{n\in\boldsymbol{Z}}).
\end{equation}
This condition ensures that the glide eigenvalues will acquire an additional negative sign. When both Condition (\ref{c1}) and Condition (\ref{c2}) are satisfied, it ensures that two energy bands with opposite eigenvalues must exchange positions and resulting in an odd number of band crossings during their evolution from point $\boldsymbol{k}$ to $\boldsymbol{k^{'}}$. Importantly, given the arbitrary choice of point $\boldsymbol{k}$ and the path of evolution, these band crossings are not isolated but form a HENL.

Specifically, let us consider a system with an $\left\{M_{z}|\frac{1}{2}0\right\}$ operation. At any arbitrary point $\boldsymbol{k}=(k_{x},k_{y})$ in the 2D BZ [see Fig.~\ref{fig2}(a)], the paired energy bands host eigenvalues of $(i)^s e^{ik_{x}/2}$ and $-(i)^s e^{ik_{x}/2}$ ($s = 0, 1$). Then, we select an endpoint $\boldsymbol{k^{'}}=(k_{x}+2\pi, k_{y})$ that satisfies conditions (\ref{c1}) and (\ref{c2}). At this point, the glide eigenvalues of the paired energy bands at $\boldsymbol{k^{'}}$ have reversed signs, becoming $-(i)^s e^{ik_{x}/2}$ and $(i)^s e^{ik_{x}/2}$. We then consider a curve $C$ which connects $\boldsymbol{k}$ and $\boldsymbol{k^{'}}$ as shown in Fig.~\ref{fig2}(a). Along the curve $C$, the typical band structure is shown in Fig.~\ref{fig2}(d). Here, because the glide eigenvalue for each band need to maintain the same sign along $C$, there should be a band crossing at an intermediate point $\boldsymbol{k}_{c}$ on the curve $C$, where the glide eigenvalues are exchanged. This band crossing is protected from gapping because of the difference in the glide eigenvalues. The choice of the path connecting $\boldsymbol{k}$ to $\boldsymbol{k^{'}}$ is arbitrary; for instance, we could select another curve $C^{'}$ as shown in Fig.~\ref{fig2}(a), which would also have an band crossing point. Thus, the crossing point at $\boldsymbol{k}_{c}$ is not isolated; the collection of such points forms a HENL, represented by the red solid curve in Fig.~\ref{fig2}(a). We note that, because of the arbitrary selection of the paired momentum $\left(\boldsymbol{k}, \boldsymbol{k^{'}}\right)$, such nodal lines will invariably traverse the BZ. The analysis for the other two glide operations, namely $\left\{M_{z}|0\frac{1}{2}\right\}$ and $\left\{M_{z}|\frac{1}{2}\frac{1}{2}\right\}$, proceeds in a similar way. The minute distinction in the selection of the terminal point $\boldsymbol{k^{'}}$ stem from the requirement that $\boldsymbol{q}$ should satisfy conditions (\ref{c1}) and (\ref{c2}). To be specific, for systems with a $\left\{M_{z}|0\frac{1}{2}\right\}$ operation, $\boldsymbol{k^{'}}$ can be chosen as $\boldsymbol{k^{'}}=(k_{x},k_{y}+2\pi)$ [the selection is not unique], as shown in Fig.~\ref{fig2}(b), and the corresponding schematic band and typical band crossings along path $C$ is depicted in Fig.~\ref{fig2}(e). Moreover, for the $\left\{M_{z}|\frac{1}{2}\frac{1}{2}\right\}$ invariant  system, the selection of $\boldsymbol{k^{'}}$-points can be two specific ones, namely $(k_{x}+2\pi, k_{y})$ and $(k_{x}, k_{y}+2\pi)$ [see Fig.~\ref{fig2}(c)]. The typical band features characteristic of HENLs can also be observed in Fig.~\ref{fig2}(f). It should be noted that the aforementioned analysis applies to both spinless and spinful systems. However, the degeneracy of the HENLs discussed here is twofold, which implies that for spinful systems, there must be a breaking of $\mathcal{PT}$ [or $\mathcal{P}\left\{\mathcal{T}|\mathbf{\boldsymbol{\tau_{1/2}}}\right\}$ ] symmetry; otherwise, the bands would be doubly degenerate everywhere.

$\widetilde{\mathcal{M}}_{z}$ represents the minimal symmetry required for the formation of HENLs, but it does not constrain the geometric configuration of HENLs within the 2D BZ. When considering additional symmetries, such as time-reversal symmetry $\mathcal{T}$ [or $\left\{\mathcal{T}|\mathbf{\boldsymbol{\tau_{1/2}}}\right\}$] , HENLs can be pinned to certain time-reversal invariant momenta (TRIM). For example, when we consider a spinless system with both $\left\{M_{z}|\frac{1}{2}0\right\}$ and $\mathcal{T}$ symmetries, i.e., a non-magnetic system, Of course, HENLs are inevitably present. Furthermore, the combined operation $\left[\ensuremath{\left\{M_{z}|\frac{1}{2}0\right\} }\ensuremath{\mathcal{T}}\right]^{2}=e^{ik_{x}}$, and at the two TRIM points, $\left(\pi,0\right)$ and $\left(\pi,\pi\right)$, $\left[\ensuremath{\left\{ M_{z}|\frac{1}{2}0\right\} }\ensuremath{\mathcal{T}}\right]^{2}=-1$. Consequently, the two bands with glide eigenvalues of $i$ and $-i$ have a Kramer-like double degeneracy, and the HENLs pass through these two TRIM, as illustrated in Fig.~\ref{fig2}(g) at points $X$ and $M$.
Similar analyses apply to other glide operations and scenarios that take into account SOC.

In addition to glide mirror $\widetilde{\mathcal{M}}_{z}$ operations that can enforce the formation of HENLs, we have discovered that in spinful systems, ordinary mirror $\mathcal{M}_{z}$ and $\mathcal{T}$ [or $\left\{\mathcal{T}|\mathbf{\boldsymbol{\tau_{1/2}}}\right\}$] can also guarantee the emergence of HENLs. The symmetry analysis follows a similar way to the previous one, considering any arbitrary point $\boldsymbol{k}$ in the 2D BZ, where the pairs of band branches can be labeled by eigenvalues $i$ and $-i$ of $\widetilde{\mathcal{M}}_{z}$, and examining the band evolution along any arbitrary path $C$ from $\boldsymbol{k}$ to $\boldsymbol{k^{'}}$ point. Here, the point $\boldsymbol{k^{'}}$ is selected as the time-reversal counterpart of $\boldsymbol{k}$, denoted as 
$\boldsymbol{k^{'}}=\mathcal{T}\boldsymbol{k}+n\boldsymbol{G}$ ($n\in\boldsymbol{Z}$), as shown in Fig.~\ref{fig2}(h). Along the path $C$, the typical band structure is shown in Fig.~\ref{fig2}(i), and one finds that, unlike $\widetilde{\mathcal{M}}_{z}$ operation, the eigenvalues of $\mathcal{M}_{z}$ for each band remains constant. And owing to the effect of time-reversal symmetry $\mathcal{T}$ [or $\left\{\mathcal{T}|\mathbf{\boldsymbol{\tau_{1/2}}}\right\}$], the energies of band at $\boldsymbol{k}$ and $\boldsymbol{k^{'}}$ are same,  yet the eigenvalues of $\mathcal{M}_{z}$ are opposite, leading to band crossings along the path $C$ [see Fig.~\ref{fig2}(i)]. The collections of these band crossings also give rise to the HENLs, as schematically shown by the red solid curve in Fig.~\ref{fig2}(h). 
\begin{table*}[t]
\centering{}%
\caption{Magnetic layer groups allowing for spinful HENL. It should be noted that  $\mathcal{PT}$ or $\mathcal{P}\left\{\mathcal{T}|\mathbf{\boldsymbol{\tau_{1/2}}}\right\}$ are forbidden in spinful systems owing to degeneracy for all the identified HENLs being twofold.}
    \begin{tabularx}{\linewidth}{llll}
    \toprule
    \toprule
    Magnetism & Minimal symmetry & Forbidden symmetry   & \multicolumn{1}{c}{  Magnetic types and candidate MLGs} \\
    \midrule
    Magnetic & $\left\{ M_{z}|\frac{1}{2}0\right\}$ & $\mathcal{PT}$, $\mathcal{P}\left\{\mathcal{T}|\mathbf{\boldsymbol{\tau_{1/2}}}\right\}$  &  Type-I: \hspace{0.15cm} 5.1.17, 7.1.28, 31.1.188, 33.1.202, 38.1.243, 41.1.276,  \\
          &       &    & \hspace{1.3cm} 43.1.298, 45.1.314 \\
          &       &    & Type-III: 31.4.191, 33.4.205, 38.5.247, 41.4.279, 43.6.303, 45.6.319  \\
          &       &    & Type-IV: 4.5.16, 5.4.20, 27.9.162, 29.6.179, 31.7.194, 31.9.196,  \\
          &       &    & \hspace{1.3cm} 36.6.226, 36.9.229 \\
          & $\left\{ M_{z}|0\frac{1}{2}\right\}$, & $\mathcal{PT}$, $\mathcal{P}\left\{\mathcal{T}|\mathbf{\boldsymbol{\tau_{1/2}}}\right\}$ & Type-I: \hspace{0.15cm} 28.1.167, 30.1.181 \\
          &       &     & Type-III: 28.4.170, 30.4.184 \\
          &       &     & Type-IV: 27.6.159, 27.8.161, 28.7.173, 30.7.187, 36.7.227, 36.8.228 \\
          & $\left\{ M_{z}|\frac{1}{2}\frac{1}{2}\right\}$, & $\mathcal{PT}$, $\mathcal{P}\left\{\mathcal{T}|\mathbf{\boldsymbol{\tau_{1/2}}}\right\}$ & Type-I: \hspace{0.15cm} 32.1.197, 34.1.207, 36.1.221, 39.1.256, 42.1.289, 46.1.323,  \\
          &       &     & \hspace{1.3cm} 48.1.343, 52.1.369, 62.1.435, 64.1.453  \\
          &       &     & Type-III: 32.4.200, 34.5.211, 36.5.225, 39.4.259, 42.5.293, 46.4.326,    \\
          &       &    &  \hspace{1.3cm} 48.4.346, 52.5.373, 62.4.438, 62.8.442,  62.9.443, 64.4.456,   \\
          &       &    &  \hspace{1.3cm} 64.8.460, 64.9.461 \\
           &       &   & Type-IV: 27.11.164, 28.6.172, 30.6.186, 31.6.193, 31.8.195, 35.7.218,    \\
           &      &    & \hspace{1.3cm} 35.8.219 \\
          & $\left\{ M_{z}|00\right\}$, $\left\{\mathcal{T}|\mathbf{\boldsymbol{\tau_{1/2}}}\right\}$ & $\mathcal{PT}$, $\mathcal{P}\left\{\mathcal{T}|\mathbf{\boldsymbol{\tau_{1/2}}}\right\}$ & Type-IV: 4.4.15, 27.7.160, 27.10.163, 27.12.165, 27.13.166, 29.7.180, \\
          &       &    & \hspace{1.3cm} 35.6.217, 35.9.220 \\
          \midrule
    Non-magnetic & $\left\{ M_{z}|\frac{1}{2}0\right\}$ & $\mathcal{P}$ &  Type-II: \hspace{0.02cm} 5.2.18, 31,2,189, 33.2.203 \\
          & $\left\{ M_{z}|0\frac{1}{2}\right\}$ & $\mathcal{P}$ &  Type-II: \hspace{0.02cm} 28.2.168, 30.2.182 \\
          & $\left\{ M_{z}|\frac{1}{2}\frac{1}{2}\right\}$ & $\mathcal{P}$ &  Type-II: \hspace{0.02cm} 32.2.198, 34.2.208, 36.2.222 \\
          & $\left\{ M_{z}|00\right\}$, $\mathcal{T}$ & $\mathcal{P}$ &  Type-II: \hspace{0.02cm} 4.2.13, 27.2.155, 29.2.175, 35.2.213, 74.2.493, 78.2.511,  \\
          &       &    &  \hspace{1.3cm} 79.2.516 \\
    \bottomrule
    \bottomrule
    \end{tabularx}%
  \label{tab2}%
\end{table*}%

So far, we have discussed the symmetry requirements for the emergence of HENLs. Based on these requirements, we search through all the 528 MLGs to screen out the candidates that may host our proposed HENL state. Specifically, for spinless systems, the existence of HENLs only requires the system to possess $\widetilde{\mathcal{M}}_{z}$ symmetry. Therefore, we have searched for MLGs containing $\widetilde{\mathcal{M}}_{z}$ operation and further categorized the candidate MLGs into two major classes: magnetic systems (type-I, type-III, and type-IV MLGs) and non-magnetic systems (type-II MLGs). The results are presented in Table~\ref{tab1}, in which we list the minimal symmetry, candidate MLGs and their corresponding types. In the case of spinful systems, the existence of HENLs requires the breaking of $\mathcal{PT}$ or $\mathcal{P}\left\{\mathcal{T}|\mathbf{\boldsymbol{\tau_{1/2}}}\right\}$ symmetry (hereafter referred to collectively as $\mathcal{PT}$), and the presence of two types of symmetries can enforce their existence, namely, $\widetilde{\mathcal{M}}_{z}$, as well as the set of $\mathcal{M}_{z}$ and $\mathcal{T}$ [or $\left\{\mathcal{T}|\mathbf{\boldsymbol{\tau_{1/2}}}\right\}$, hereafter referred to collectively as $\mathcal{T}$)]. Accordingly, in the MLGs that lack $\mathcal{PT}$ symmetry, we have identified for candidate groups with $\widetilde{\mathcal{M}}_{z}$ operations and also for those with both $\mathcal{M}_{z}$ and $\mathcal{T}$ operations. The search results are presented in Table~\ref{tab2}
. From the two tables, we can observe that HENLs enforced by $\widetilde{\mathcal{M}}_{z}$ symmetry can exist in all four types of MLGs for both spinless and spinful systems. In contrast, HENLs enforced by $\mathcal{M}_{z}$ and $\mathcal{T}$ symmetry can only exist in spinful type-II and type-IV MLGs. Furthermore, the candidate magnetic group numbers and their corresponding group symbols are also presented in Table~\ref{tab3}. Our tables cover previous study~\cite{damljanovic2023movable}, which identified some unavoidable NLs in non-magnetic systems. Moreover, our study is based on a general symmetry analysis, thereby allowing the distribution of HENLs to encompass 528 magnetic space groups, with broader universality and richer content. It should be clarified that some of the candidate groups in the list have HENLs pinned to HSLs due to additional crystalline symmetries. Breaking the corresponding symmetries allows these HENLs to deviate from the high-symmetry paths.

In addition, it is worth noting that when materials meet specific electronic filling conditions, HENLs invariably appear near the Fermi surface. This is a consequence of the fact that the bands of HENLs always emerge in pairs. Specifically, in spinless systems, the 4$N$+2 fillings are required, while for spinful systems, the required fillings is 2$N$+1.

\begin{table*}[t]
\centering{}%
\caption{Correspondence between magnetic layer group number and group symbol. Here, magnetic layer group number is the OG number, the group symbol is denoted by the OG symbol.}
    \begin{tabularx}{\linewidth}{lX|X}
    \toprule
    \toprule
    Type & \multicolumn{1}{c}{Magnetic layer group number}  &  \multicolumn{1}{c}{OG symbol} \\
    \midrule
     Type-I: & 5.1.17, 7.1.28, 28.1.167, 30.1.181, 31.1.188, 32.1.197, 33.1.202, 34.1.207, 36.1.221,38.1.243, 39.1.256, 41.1.276, 42.1.289, 43.1.298, 45.1.314, 46.1.323, 48.1.343, 52.1.369, 62.1.435, 64.1.453 & $p11a$,   $p112/a$, $pm2_{1}b$, $pb2b$, $pm2a$, $pm2_{1}n$, $pb2_{1}a$, $pb2n$, $cm2a$, $pmaa$, $pban$, $pmma$, $pman$, $pbaa$, $pbma$, $pmmn$, $cmme$, $p4/n$, $p4/nbm$, $p4/nmm$   \\
      Type-II: & 5.2.18, 7.2.29, 28.2.168, 30.2.182, 31.2.189, 32.2.198, 33.2.203, 34.2.208, 36.2.222, 38.2.244, 39.2.257, 41.2.277, 42.2.290, 43.2.299, 45.2.315, 46.2.324, 48.2.344, 52.2.370, 62.2.436, 64.2.454  & $p11a1'$, $p112/a1'$, $pm2_{1}b1'$, $pb2b1'$, $pm2a1'$, $pm2_{1}n1'$, $pb2_{1}a1'$, $pb2n1'$, $cm2a1'$, $pmaa1'$, $pban1'$, $pmma1'$, $pman1'$, $pbaa1'$, $pbma1'$, $pmmn1'$, $cmme1'$, $p4/n1'$, $p4/nbm1'$, $p4/nmm1'$   \\
      Type-III: & 7.3.30, 28.4.170, 30.4.184, 31.4.191, 32.3.200, 33.4.205, 34.5.211, 36.5.225, 38.5.247, 38.7.249, 38.9.251, 39.4.259, 39.7.262, 41.4.279, 41.8.283, 41.9.284, 42.5.293, 42.7.295, 42.9.297, 43.6.303, 43.7.304, 43.9.306, 45.6.319, 45.7.320, 45.8.321, 46.4.326, 46.7.329, 48.4.346, 48.7.349, 52.5.373, 62.4.438, 62.8.442, 62.9.443, 64.4.456, 64.8.460, 64.9.461  & $p112'/a$, $pm'2'_{1}b$, $pb'2'b$, $pm'2'a$, $pm'2'_{1}n$, $pb'2'_{1}a$, $pb'2'n$, $cm'2'a$, $pm'a'a$, $pm'aa$, $pma'a$, $pb'a'n$, $pba'n$, $pm'm'a$, $pmm'a$, $pm'ma$, $pm'a'n$, $pma'n$, $pm'an$, $pb'a'a$, $pb'aa$, $pba'a$, $pb'm'a$, $pbm'a$, $pb'ma$, $pm'm'n$, $pmm'n$, $cm'm'e$, $cmm'e$, $p4'/n$, $p4/nb'm'$, $p4'/nbm'$, $p4'/nb'm$, $p4/nm'm'$, $p4'/nmm'$, $p4'/nm'm$    \\
      Type-IV: & 4.5.16, 5.4.20, 6.7.27, 7.6.33, 27.6.159, 27.8.161, 27.9.162, 27.11.164, 28.6.172, 28.7.173, 29.6.179, 30.6.186, 30.7.187, 31.6.193, 31.7.194, 31.8.195, 31.9.196, 35.7.218, 35.8.219, 36.6.226, 36.7.227, 36.8.228, 36.9.229, 37.8.237, 37.10.239, 37.12.241, 38.10.252, 38.11.253, 38.12.254, 38.13.255, 40.10.272, 40.12.274, 41.10.285, 41.11.286, 41.12.287, 41.13.288, 47.9.338, 47.11.340, 47.12.341, 48.8.350, 48.9.351, 48.10.352, 48.11.353, 51.6.367, 61.10.431, 61.12.433   & $p_{2a}11m'$, $p_{2b}11a$, $p_{2a}112'/m'$, $p_{2b}112/a$, $p_{2b}m'2m'$, $p_{2b}m2'm'$, $p_{2a}m'2m'$, $p_{c}m'2m'$, $p_{2a}m2'_{1}b'$, $p_{2a}m2_{1}b$, $p_{2a}b'2_{1}m'$, $p_{2a}b'2b'$, $p_{2a}b2b$, $p_{2b}m2'a'$, $p_{2b}m'2'a$, $p_{2b}m'2a'$, $p_{2b}m2a$, $c_{p}m'2m'$, $c_{p}m2'm'$, $c_{p}m2a$, $c_{p}m'2a'$, $c_{p}m2'a'$, $c_{p}m'2'a$, $p_{2a}mm'm'$, $p_{2am'mm'}$, $p_{c}mm'm'$, $p_{2b}m'aa'$, $p_{2b}ma'a'$, $p_{2b}m'a'a$, $p_{2b}maa$, $p_{2b}m'am'$, $p_{2b}ma'm'$, $p_{2b}m'ma'$, $p_{2b}m'm'a$, $p_{2b}mm'a'$, $p_{2b}mma$, $c_{p}m'm'm'$, $c_{p}mm'm'$, $c_{p}mmm'$, $c_{p}m'm'e'$, $c_{p}m'me'$, $c_{p}mm'e'$, $c_{p}mme'$, $p_{p}4/m'$, $p_{p}4/m'm'm'$, $p_{p}4/m'mm$     \\
    \bottomrule
    \bottomrule
    \end{tabularx}%
  \label{tab3}%
\end{table*}%

\section{Lattice model}
To further confirm the existence of the HENLs, we explicitly construct minimal lattice models guided by the candidate MLGs. First, let us consider MLG 5.1.17,  listed in Table~\ref{tab1}, which only contains the operation $\left\{ M_{z}|\frac{1}{2}0\right\}$. And we can derive a two-band lattice model by setting one $\left|s\right\rangle $ orbital at 2a Wyckoff position $\left\{ \left(x,y,z|m_{x},m_{y},m_{z}\right),\left(x+\frac{1}{2},y,-z|-m_{x},-m_{y},m_{z}\right)\right\}$ with $x=\frac{3}{4}$, $y=\frac{1}{4}$, $z=0$ and $m_{x}=m_{y}=m_{z}=0.2$. In the absence of SOC, the matrix representation of the symmetry operator, under the basis of the two sites, can be represented as 
\begin{equation}\label{mzx}
\left\{ M_{z}|\frac{1}{2}0\right\}	=   \sigma_{x},
\end{equation}
with $\sigma_{i}$ ($i = x, y, z$) the Pauli matrix and $\sigma_{0}$ the $2\times2$ identity matrix. Then, the symmetry allowed model Hamiltonian may be
written as~\cite{zhang2022magnetictb,zhang2023magnetickp}
\begin{eqnarray}\label{tb1}
\mathcal{H}(\boldsymbol{k})_{5.1.17} & = & \left[t_{1}\sin\frac{k_{x}}{2}+t_{2}\cos\frac{k_{x}}{2}-r_{1}\sin\left(\frac{k_{x}}{2}-k_{y}\right)\right. \nonumber \\
 &  & \left.+r_{2}\cos\left(\frac{k_{x}}{2}-k_{y}\right)\right]\sigma_{x}+e_{1}\sigma_{0}.
\end{eqnarray}

\begin{figure}[t]
\includegraphics[width=8.4cm]{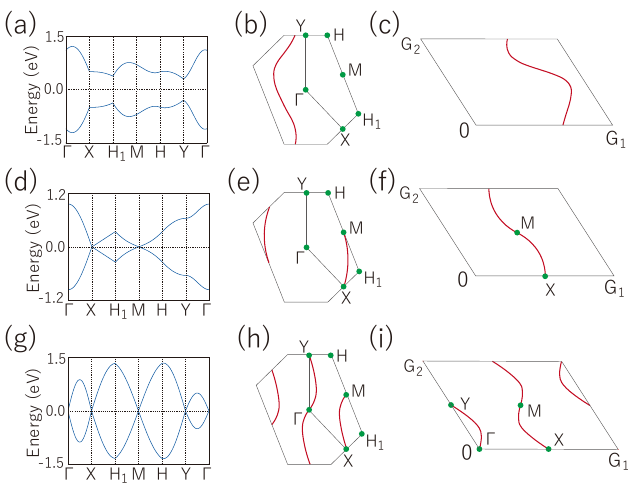}
\caption{(a), (d) and (g) show the band structures along high-symmetry lines for lattice models (\ref{tb1}), (\ref{tb2}), and (\ref{tb3}). (b), (e) and (h) display the BZs as well as HENL distributions for the same models. (c), (f) and (i) present the HENL distributions in the unit cells of the reciprocal lattices.
\label{fig3}}
\end{figure}

Here, $e_{1}$, $t_{1(2)}$ and $r_{1(2)}$ are model parameters. The band structure of this model (\ref{tb1}) with $e_{1}=0$, $t_{1}=-0.15$, $t_{2}=-0.175$, $r_{1}=-0.025$ and $r_{2}=-0.1$ is presented in Fig.~\ref{fig3}(a) [the units of the model parameters here and below are all eV]. Interestingly, this band structure seems far from the NLs semimetal and is more akin to an insulator. Then, we carefully searched for possible band crossing points within the BZ, as shown in Fig.~\ref{fig3}(b). One can observe that HENLs indeed exist within the BZ, but they are missed along the conventional high-symmetry paths of oblique BZ, which further demonstrates the \textquotedblleft hidden$\textquotedblright$  nature of HENLs. Furthermore, we demonstrate the distribution of HENLs within the unit cell of the reciprocal lattice, which can be observed to be consistent with the analysis presented in the previous section. And it is easy to verify that HENLs will always exist regardless of the choice of model parameters.

Subsequently, we check the effects of time-reversal symmetry on HENLs. We then consider a type-II MLG 5.2.18 with symmetry operator of $\left\{ M_{z}|\frac{1}{2}00\right\}$ and $\mathcal{T}$ as listed in Table~\ref{tab1}. Similarly, an s-orbital is placed at the 2a position $\left\{ \left(\frac{3}{4},\frac{1}{4},0\right),\left(\frac{1}{4},\frac{1}{4},0\right)\right\} $, and the matrix representation of the symmetry operator can be expressed as
\begin{equation}\label{mzT}
\left\{ M_{z}|\frac{1}{2}0\right\}	=   \sigma_{x}, \quad \mathcal{T}=\sigma_{0}\mathcal{K},
\end{equation}
where $\mathcal{K}$ denotes the complex conjugation operator. The spinless Hamiltonian under the above symmetries constraint in  Eq.(\ref{mzT}) may be written as
\begin{equation}\label{tb2}
\mathcal{H}(\boldsymbol{k})_{5.2.18}	=	\left[t\ \cos\frac{k_{x}}{2}+r\ \cos\left(\frac{k_{x}}{2}-k_{y}\right)\right]\sigma_{x}+e_{1}\sigma_{0}.
\end{equation}
The band structure of Eq.(\ref{tb2}) with $e_{1}=0$, $t=-0.2$, $r=-0.04$ is plotted in Fig.~\ref{fig3}(d). We can observe two band crossings at high-symmetry point $X$ and $M$. As we have mentioned, such band crossings may not be isolated, and they are part of the HENL. The distribution of HENL in BZ is shown in Fig.~\ref{fig3}(e), where we can observe that HENL indeed passes through the $X$ and $M$ points. Moreover, the shape of HENL in unit cell of reciprocal lattice is also presented in Fig.~\ref{fig3}(f), and one can verify that it always exists and is pinned by the two TRIM points $X$ and $M$ due to 
$\left[\ensuremath{\left\{M_{z}|\frac{1}{2}0\right\} }\ensuremath{\mathcal{T}}\right]^{2}=-1$ at these points.

On the other hand, as aforementioned, in addition to the glide mirror  symmetry $\widetilde{\mathcal{M}}_{z}$, the ordinary mirror $\mathcal{M}_{z}$ and time-reversal symmetry $\mathcal{T}$ also enforce the existence of HENLs in spinful systems. As an illustration, let us consider MLG 4.2.13 as listed in Table~\ref{tab2}, which only hosts operations of $\mathcal{M}_{z}$ and $\mathcal{T}$. The lattice model can be established by putting $\left\{ \left|s\uparrow\right\rangle ,\left|s\downarrow\right\rangle \right\}$ orbitals on 1a Wyckoff position $\left(0,0,0\right)$. In the presence of SOC, the matrix representation of the symmetries reads
\begin{equation}\label{MT}
M_{z}	=  -i \sigma_{z}, \quad \mathcal{T}=i \sigma_{y}\mathcal{K},
\end{equation}
Then, the spinful lattice Hamiltonian may be written as
\begin{eqnarray}\label{tb3}
\mathcal{H}(\boldsymbol{k})_{4.2.13} & = & \left[t_{1}\sin(k_{x}-k_{y})+r_{1}\sin k_{y}+r_{2}\sin k_{x}\right]\sigma_{z} \nonumber \\
 &  & +\left[e_{1}+t_{2}\cos(k_{x}-k_{y})+r_{3}\cos k_{y}\right.+ \nonumber \\
 &  & \left.+r_{4}\cos k_{x}\right]\sigma_{0}
\end{eqnarray}
In Fig.~\ref{fig3}(g), we plot the band structure for the lattice model (\ref{tb3}) with $r_{1}=0.1$, $r_{2} = 0.175$ and other parameters are set to 0. We observe the Kramers degeneracy at the four TRIM: $\Gamma$, $X$, $M$ and $Y$, arising from $\mathcal{T}^{2}=-1$. Similar to the previous case, we have searched for and confirmed the distribution of HENLs in the BZ and unit cell of reciprocal lattice, as shown in Figs.~\ref{fig3}(h) and 3(i). The existence of these HENLs remains robust against parameter variations, exhibiting essential characteristics. Before closing this section, we would like to point out that, in addition to $\mathcal{T}$, other crystal symmetries also constrain the distribution of HENLs in the BZ. We shall demonstrate that in real materials, HENLs can exhibit a rich pattern.

\begin{figure}[t]
\includegraphics[width=8.1cm]{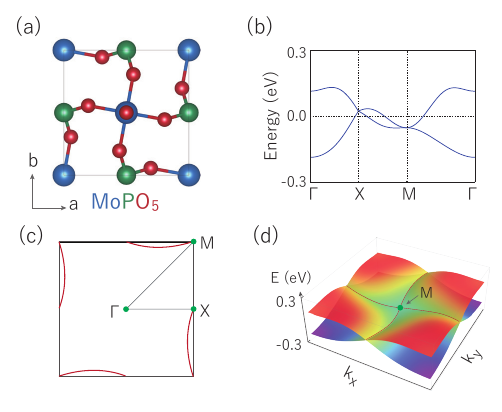}
\caption{(a) Stucture and (c) BZ of monolayer MoPO$_5$. (b) The band structure of MoPO$_5$ without SOC. (d) 3D band plotting of HENL.
\label{fig4}}
\end{figure}
\begin{figure}[t]
\includegraphics[width=8.4cm]{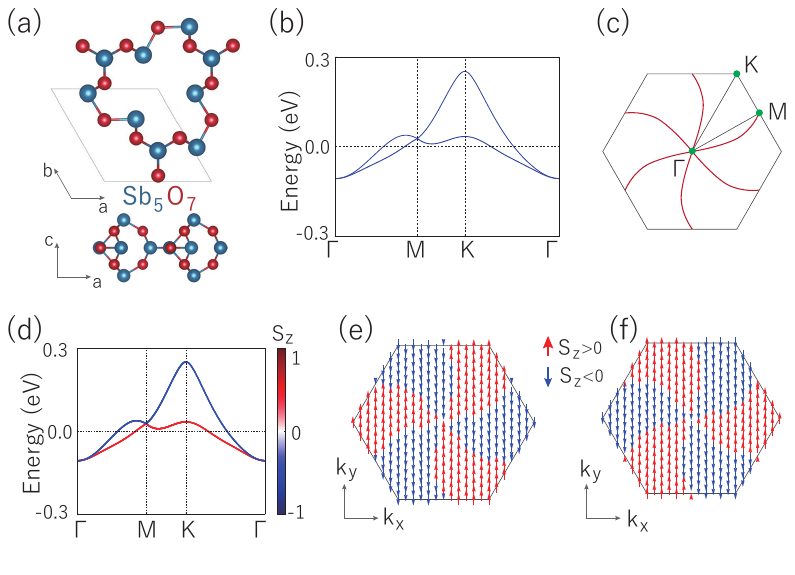}
\caption{(a) Top and side view of monolayer Sb$_{5}$O$_7$ and (c) corresponding to BZ with the distribution of HENLs. (b) Band structure of Sb$_{5}$O$_7$ with SOC. (d) Spin-resolved band structures along high-symmetry lines. The spin configurations of $S_{z}$ for (e) lower-band and (f) upper-band within the BZ.
\label{fig5}}
\end{figure}

\section{Spinless material example: MoPO$_5$}
To demonstrate that HENLs can indeed exist in real materials, we have searched the 2D materials with the target MLGs and identified two examples. The first example is molybdenum phosphate MoPO$_5$~\cite{kierkegaard1964crystal, lezama1993reduced}, and its primitive cell of monolayer is shown in Fig.~\ref{fig4}(a). The  paramagnetic MoPO$_5$ has a tetragonal structure with MLG 52.2.370 (i.e., the layer group p4/n), which is one candidate in Table~\ref{tab1}. The MLG 52.2.370 is generated by a two-fold screw-rotation $\widetilde{C_{2z}}=\left\{ C_{2z}|\frac{1}{2}\frac{1}{2}\right\}$, a four-fold screw-rotation $\widetilde{C_{4z}}=\left\{ C_{4z}|\frac{1}{2}0\right\}$ and a spatial inversion $\mathcal{P}$. Combing $\widetilde{C_{2z}}$ and $\mathcal{P}$ leads to a glide mirror operation $\left\{M_{z}|\frac{1}{2}\frac{1}{2}\right\}$, which is a critical operation for the formation of HENLs. In addition, we focus on the paramagnetic phase of the material, so the $\mathcal{T}$ is also preserved. 

We performed first-principles calculations using the Vienna ab initio simulation package (VASP)~\cite{PhysRevB.49.14251,PhysRevB.54.11169,PhysRevB.50.17953} with the Perdew-Burke-Ernzerhof (PBE) type exchange-correlation~\cite{PhysRevLett.77.3865}. A 600 eV cutoff energy and a $9\times9\times1$ $\Gamma$-centered $k$-mesh were used, with convergence criteria of $10^{-7}$ eV for energy and 0.01 $\textrm{eV}/\mathring{\textrm{A}}$ for forces.
The band structure of MoPO$_5$ without SOC is shown in Fig.~\ref{fig4}(b). It is clearly observed that the bands near the Fermi level are well separated from other bands, forming two ideal isolated bands. Such a situation is consistent with the electronic filling principle of spinless HENLs we have previously discussed, as the primitive cell of the MoPO$_5$ (Mo: $4s^{2}4p^{6}4d^{5}5s^{1}$, P: $3s^{2}3p^{3}$, O: $2s^{2}2p^{4}$) has 98 valence electrons, satisfying the 4$N$+2 rule (where $N$=24). 

Furthermore, one observes that there exist two band crossings at $X$ and $M$ points, and an another band crossing is present at the $Y$ $(0,\pi)$ point due to the $\widetilde{C_{4z}}$ symmetry. Similar to the previous analysis, the Kramers-like degeneracy at these three TRIMs originates from the combined operation $\left\{M_{z}|\frac{1}{2}\frac{1}{2}\right\}\mathcal{T}$, which satisfies $\left[\ensuremath{\left\{M_{z}|\frac{1}{2}\frac{1}{2}\right\} }\ensuremath{\mathcal{T}}\right]^{2}=-1$. In fact, these three points are not isolated due to the presence of $\left\{M_{z}|\frac{1}{2}\frac{1}{2}\right\}$. A careful scan shows
that they are located on a HENL in the 2D BZ, as schematically shown in Fig.~\ref{fig4}(c), consistent with the above symmetry analysis. Interestingly, we plot the band structure centered at the $M$ point in Fig.~\ref{fig4}(d), we can observe that, due to the presence of $\widetilde{C_{4z}}$ symmetry, HENLs exhibit a novel  fourfold fan-like pattern.

\section{Spinful	material example: Sb$_{5}$O$_7$}
We further motivate our investigation by considering a second material example. It is the monolayer compound Sb$_{5}$O$_7$~\cite{tang2024screening}. The Sb$_{5}$O$_7$ belongs to MLG 74.2.493 (layer group $\textrm{p}\overline{6}$), listed in Table~\ref{tab2}, and the primitive cell and side view of crystal are presented in Fig.~\ref{fig5}(a). The MLG 74.2.493 contains a mirror $\mathcal{M}_{z}$, a threefold roto-reflection $S_{3z}$, and time-reversal symmetry $\mathcal{T}$.

The calculated band structure for Sb$_{5}$O$_7$ with SOC is plotted in Fig.~\ref{fig5}(b). The parameter settings match prior settings and have converged well. Importantly, there are several expected band features near the Fermi level. Firstly, the band structure is remarkably ideal, with only one set of bands crossing the Fermi level, facilitating experimental observation. We note that, the valence electron number of Sb$_{5}$O$_7$ (Sb: $5s^{2}5p^{3}$, O: $2s^{2}2p^{4}$) is 67, which satisfies the filling criterion for spinful HENLs with the 67 = 2$N$ + 1 (where $N$ = 33). Secondly, because of the breaking of inversion symmetry ($\mathcal{P}$) in the system, spin degeneracy is not permitted except at the four TRIM points. As momentum moves away from the four TRIM points, the bands exhibit significant spin splitting, attributed to the strong SOC of Sb atom. According to Table~\ref{tab2}, MLG 74.2.493 may host the HENLs owing to presence of $\mathcal{M}_{z}$ and $\mathcal{T}$. We then perform a  careful scan and result is shown in Fig.~\ref{fig5}(c). One observes that there indeed exist HENLs in the BZ, crossing four TRIM points. And  the HENLs here exhibit a sixfold fan-like pattern owing to the presence of $C_{3}\mathcal{T}$ symmetry.

Additionally, it is worth investigating the spin-related properties of HENLs as a result of the system lacking inversion symmetry. We then plot the spin-resolved band structure as shown in Fig.~\ref{fig5}(d). One can see that the composition of HENLs originates from two spin-polarized branches of bands. The spin of these states is oriented along $z$ with only an $S_{z}$ component, owing to presence of $\mathcal{M}_{z}$ symmetry and the fact that forming two band being well separated from the other bands. Next, we explore the distribution of $S_{z}$ in BZ. The spin texture for lower-band and upper-band are presented in Fig.~\ref{fig5}(e) and (f), respectively. Notably, we observe a uniform spin structure in the gap region of the BZ, where a momentum-independent spin configuration is present, known as the persistent spin texture (PST)~\cite{RevModPhys.89.011001, tao2018persistent}. We note that the lower-band and the upper-band exhibit opposite PST. Moreover, the PST changes abruptly at HENLs, where the $S_{z}$ component of the spin is reversed. This result means that the PST can be effectively modulated by HENLs.

\section{CONCLUSIONS AND DISCUSSION}
In summary, we propose a new class of semimetal phase in two dimensions: the HENL phase. Such NLs are not easily detectable, yet they are ubiquitous in solids, as their existence only requires the support of a single glide mirror symmetry. HENLs can be realized in 200 of the 528 MLGs, including both spinless and spinful systems. As discussed, when the electron filling satisfies the condition of 4$N$+2 in spinless system (2$N$+1 for spinful systems), the presence of the HENLs in candidate MLGs is required regardless of the details of the systems. The distribution of HENLs enforced by the glide mirror in the BZ is completely irregular. The presence of time-reversal symmetry allows HENLs to pass through high-symmetry points, making them slightly more observable. In addition, other crystalline symmetries, such as rotation, screw-rotation and vertical mirror, may further constrain the shape of HENLs, and even cause them to appear on high-symmetry lines. Furthermore, we have predicted two material candidates, MoPO$_5$ and Sb$_{5}$O$_7$, which respectively host spinless and spinful HENLs. Interestingly, we have revealed that spinful HENLs exhibit properties of PST distinct from those of traditional NLs. The regions of PSTs are separated by HENLs acting as domain walls, thus allowing for their manipulation by HENLs. Such novel spin-related phenomena may provide a new playground for the application of topological physics in spintronics. The concept of HENLs can be readily extended to three dimensions and is applicable across a spectrum of systems, including electronic, phononic, and various artificial systems.

\begin{acknowledgments}
This work is supported by the National Natural Science Foundation of China (Grant No. 12304086, Grant No. 12304165, No. 12234003, No. 12321004), the Natural Science Foundation of Inner Mongolia Autonomous Region (Grants No. 2023QN01003), the Startup Project of Inner Mongolia University (Grant No. 21200-5223730).
\end{acknowledgments}
\bibliography{ref}
\end{document}